\documentclass[]{article}
\usepackage[utf8]{inputenc}
\usepackage[T1]{fontenc}

\usepackage[fleqn]{amsmath}
\usepackage{verbatim,amsthm}
\usepackage{graphicx}
\usepackage[colorinlistoftodos]{todonotes}
\usepackage[colorlinks=true, allcolors=blue]{hyperref}
\usepackage{amssymb,xcolor,amscd}
\usepackage{float}
\usepackage{enumitem}
\usepackage{listings}
\usepackage{authblk}
\usepackage{dsfont}

\newcommand{\Nd}{\mathbb{N}}

\newcommand{\fy}{\varphi}

\newcommand{\imply}{\Rightarrow}             

\let\0=\varnothing

\DeclareMathOperator{\Mean}{E}

\renewcommand{\d}{\mathrm{d}}

\newcommand{\e}{\mathrm{e}}                  

\theoremstyle{plain}

\newtheorem{prop}{\bfseries Proposition.}[section]

\newtheorem{rem}{\scshape\mdseries Remark}[section]

\numberwithin{equation}{section}

\title{On the Generic Cut--Point Detection Procedure in the Binomial Group Testing}

\author[1]{Ugnė Čižikovienė}
\author[1]{Viktor Skorniakov\thanks{corresponding author; e-mail: viktor.skorniakov@mif.vu.lt}}
\affil[1]{Institute of Applied Mathematics, Faculty of Mathematics and Informatics, Vilnius University, Naugarduko 24, Vilnius LT-03225, Lithuania}

\date{}

\begin{document}

\maketitle

\begin{abstract}
    Initially announced by Dorfman in 1943, (Binomial) Group Testing (BGT) was quickly recognized as a useful tool in many other fields as well. To apply any particular BGT procedure effectively, one first of all needs to know an important operating characteristic, the so called Optimal Cut-Point (OCP), describing the limits of its applicability. The determination of the latter is often a complicated task. In this work, we provide a generic algorithm suitable for a wide class of the BGT procedures and demonstrate its applicability by example. The way we do it exhibits independent interest since we link the BGT to seemingly unrelated field --- the bifurcation theory.
\end{abstract}

\section{Introduction}\label{s:intro}

Group Testing (GT) refers to a set of methods for finding defectives in a given cohort of items. The first and likely the best known GT procedure was suggested by Dorfman \cite{dorfman_detection_1943}. During World War II, he sought for a cost-saving screening for Syphilis of US soldiers and proposed the following method. Given $n$ blood samples, pool them and test. In case the pooled sample tests negative, finish; in case it tests positive, retest each individual.

Since appearance of that paper, many GT procedures were proposed. As the name suggests, they all share one feature: in the process of testing (at least in some stages), items are grouped and tested in pools. In other respects there can be significant differences. It is common practice to distinguish between Probabilistic Group Testing (PGT) and Combinatorial Group Testing (CGT). In PGT, one assumes that each tested item can be defective with some probability; in CGT, one assumes that in the tested cohort there are $d$ defectives and the defectiveness generating mechanism is not considered. For a good review of some of the PGT procedures, we refer to Malinovsky \cite{malinovsky_revisiting_2019}; Aldridge et. al. \cite{aldridge2019group} covers CGT and provides some other classification criteria as well.

In this paper, we focus on the subclass of PGT called the Binomial Group Testing (BGT). In BGT, one restricts PGT by further assuming that:
\begin{itemize}
    \item[(BGT1)] each item is independent of the rest;
    \item[(BGT2)] the probability of defectiveness $p\in(0,1)$ is the same for all items in the tested cohort.
\end{itemize}
An intuition suggests that PGT makes sense only when the defectiveness is rare. In the frame of BGT, this concept is formalized as follows. Given BGT procedure $X$, denote:
\begin{itemize}
    \item $T_{X}(n,p)=$"the random number of tests utilized per cohort spanning $n$ items";
    \item $M_X(n,p)=\Mean T_{X}(n,p)=$"average number of tests utilized per cohort spanning $n$ items";
    \item $t_{X}(n,p)=\frac{M_{X}(n,p)}{n}=$"number of tests spent per single item on average when testing a cohort spanning $n$ items".
\end{itemize}
One says that, for a fixed $p$, procedure $X$ makes sense provided $\inf_{n\in\Nd}t_{X}(n,p)<1$. For example, in the above described Dorfman procedure $D$, one often ends up with a single test whenever the prevalence of the disease $p$ is small; assuming an opposite (i.e., large $p$), $M_{D}(n,p)\approx n+1$ this way making procedure insensible. The same logic applies to other procedures as well. Therefore, one can expect that, for large $p$'s, any BGT procedure makes no sense. This indeed holds true. Namely, there is a fundamental result due to Ungar \cite{Ungar-1960}. It says that, for $p> \frac{3-\sqrt{5}}{2}$, $\inf_{n\in\Nd} t_X(n,p)>1$ for any BGT. In words, under BGT assumptions (BGT1)--(BGT2), it is optimal to test items individually and do not invoke GT if $p> \frac{3-\sqrt{5}}{2}$. On the other hand, in the same paper, it was shown that, for $p<\frac{3-\sqrt{5}}{2}$, individual testing is not optimal and there always exists a BGT procedure with $\inf_{n\in\Nd}t_X(n,p)<1$. In the GT literature, the value $\frac{3-\sqrt{5}}{2}$ is termed the Universal Cut-Point (UCP). However, it is often the case that particular procedure has its own Optimal Cut-Point (OCP) $p_{X,c}<\mathrm{UCP}$. The detection of this operational characteristic is an important and usually not a straightforward task. In this paper, we describe a generic algorithm for finding of $p_{X,c}$ suitable for a wide class of BGT procedures. 

The remaining part of the paper contains two sections and two appendixes. In Section \ref{s:results}, we state our main result. In Section \ref{s:examples-discussion}, we provide examples of applications and a short accompanying discussion. In appendixes, we place proofs and figures.

\section{Results}\label{s:results}

\subsection{Notions and Assumptions}\label{ss:notions_assumptions}
In the remaining part of the paper, $X$ stands for the unspecified (generic) BGT procedure. Corresponding quantities are sub-scripted: $M_{X}(n,p),t_{X}(n,p)$, etc. When turning to the particular procedure, $X$ is replaced by another letter. $\mathrm{UCP}:=\frac{3-\sqrt{5}}{2}$ stands for the previously mentioned Universal Cut-Point. For some procedures, $n\in\Nd$ is not an actual size of the tested cohort $N(n)$ but a parameter. For example, in case of procedure described in Subsection \ref{ss:p_A2}, the size of the tested cohort $N(n)=n^2$ for $n\in\{2,3,\ldots\}$. Such parametrization redefines $t_X(n,p)$ making it equal to $\frac{M_X(n,p)}{N(n)}$ yet it does not affect our results. Therefore, we retain all previous notions unchanged. When dealing with a particular procedure, the role of $n$ is unambiguous.

In what follows, we focus on the subset of the BGT procedures with a mean $M_X(n,p)$ satisfying assumptions listed below.
\begin{itemize}
    \item[(M0)] $\exists c\geq 2$ s.t. $X$ is a-priori known to be useless for $n\in[1,c)$.
    \item[(M1)] Function $\Nd\times (0,\mathrm{UCP}]\ni (n,p) \mapsto M_X(n,p)$ can be treated as a continuous function on $[c,\infty)\times (0,\mathrm{UCP}]$ differentiable in the whole interior of its domain.
    \item[(M2)] $\forall n\in(c,\infty)$ function $(0,\mathrm{UCP}]\ni p\mapsto M_X(n,p)$ is strictly increasing.
    \item[(M3)] $\forall n\in (c,\infty)\, t_X(n,\mathrm{UCP})>1$.
    \item[(M4)] $\forall n\in (c,\infty)\,\exists\, p\in(0,\mathrm{UCP}): t_X(n,p)<1$.
\end{itemize}
(M0) may be viewed as a condition required to restrict the range of the dynamical system we make use of when looking for the OCP $p_{X,c}$. Together with (M3), it also defines the boundary value for determination of $p_{X,c}$ for the case $p_{X,c}=\mathrm{UCP}$ (see examples and discussion in Section \ref{s:examples-discussion}). In a usual case, one can set $c=2$ and have it since, in this context, $c=2$ means that taking one item we do not need any GT procedure: to identify the defectiveness of a single item, one always needs one test.

Constraint (M1) is the most restrictive since not all BGT procedures can be naturally extended to have a differentiable mean with respect to both arguments (in case of $p$ this always holds true). However, this particular assumption is the one we rely on. It also enables us to draw the connection with the bifurcation theory.

Other constraints can be justified naturally and attributed to many of the BGT procedures in general.

(M2) states that an average number of tests per batch spanning $n$ items should increase together with the rate of defectiveness. For justification, we mention another fundamental result due to Yao and Hwang \cite{Yao-monotonicity-1988} who have demonstrated that, $\forall n\in\Nd$, function $ (0,\mathrm{UCP}]\ni p\mapsto \inf_X M_X(n,p)$ is strictly increasing (with an infimum being taken over all possible BGT procedures).

Finally, (M4) in technical terms states that we focus on the procedures applicable to any number of tested items at least for some $p$'s in the range of their sensibility. This is very often the case since many procedures are suitable for large scale testing when the rate of defectiveness is small.

\subsection{Statement}\label{ss:statement}
Our results are given in two propositions stated below. The first one characterizes properties of the OCP. 

\begin{prop}\label{prop:ocp}
Assume (M0)--(M4). Let $p_{X,c}=\sup\{p\in(0,\mathrm{UCP})\mid \exists n\in(c,\infty): t_X(n,p)<1\}$. Then $\forall p\in(0,p_{X,c})$ procedure $X$ makes sense on the continuous scale; $\forall p\in(p_{X,c},\mathrm{UCP}]$ it makes no sense at all, that is, 
\begin{equation}\label{e:makes_no_sense}
    (n,p)\in (c,\infty)\times (p_{X,c},\mathrm{UCP}]\imply t_{X}(n,p)>1.    
\end{equation}
\end{prop}
The second proposition demonstrates that, under (M0)--(M4), there exists a generic procedure for finding $p_{X,c}$, and it can be naturally cast in terms of the bifurcation theory as follows. Treating $p\in(0,\mathrm{UCP}]$ as a control parameter and $n\in (c,\infty)$ as a function of some latent continuous argument, consider autonomous dynamical system
\begin{equation}\label{e:ds1}
    \Dot{n}=t_X(n,p)-1.
\end{equation}
\begin{prop}\label{prop:bifurcations}
Assume (M0)--(M4). $p_{X,c}$ is a bifurcation point of the system \eqref{e:ds1} and one can distinguish between three types of possible bifurcations.
\begin{itemize}
    \item[(b0)] $p_{X,c}$ is the only value of the control parameter for which \eqref{e:ds1} admits fixed points in $(c,\infty)$. In this case $p_{X,c}<\mathrm{UCP}$ and all $n\in(c,\infty)$ solve $t_X(n,p_{X,c})=1$.
\end{itemize}
If there exists $p_l\in(0,p_{X,c})$ for which \eqref{e:ds1} admits a fixed point $n\in(c,\infty)$, there are two possibilities:
    \begin{itemize}
        \item[(b1)] \eqref{e:ds1} has fixed points in $(c,\infty)$ for all $p\in[p_l,p_{X,c})$ yet there are no fixed points corresponding to $p_{X,c}$;
        \item[(b2)] \eqref{e:ds1} has fixed points in $(c,\infty)$ for all $p\in[p_l,p_{X,c}]$ including $p_{X,c}$ which then is necessary smaller than UCP.
    \end{itemize}
In all cases, bifurcation curve induces a differentiable map $(c,\infty)\ni n\mapsto p_n\in (0,p_{X,c}]$. Therefore, $p_{X,c}$ can be determined by finding its maximum. For bifurcations of types (b0) and (b2), this amounts to solving a system 
\begin{equation}\label{e:p_Xc_system}
    \begin{cases}
        &t_X(n,p)=1,\\
        &\frac{\partial}{\partial n}t_{X}(n,p)=0
    \end{cases}
\end{equation}
(with respect to both $n$ and $p$) and then picking up a largest $p$ value from the set $S=\{(n,p)\in(c,\infty)\times (0,\mathrm{UCP}]\mid (n,p)$ solves \eqref{e:ds1}$\}$. For the bifurcation of type (b1), $p_{X,c}=\max(\lim_{n\to c+}p_n,\lim_{n\to\infty}p_n)$. In particular, this holds true when \eqref{e:p_Xc_system} has no solution lying in $(c,\infty)\times(0,\mathrm{UCP}]$.
\end{prop}
Before proceeding to examples, we provide several remarks.
\begin{rem}\label{r:about_scale}
Prop. \ref{prop:bifurcations} establishes the procedure for finding OCP on the continuous scale (COCP). In practice, one operates on the discrete one since the number of tested items $n$ is integer. As a rule, discrete scale OCP (DOCP) is lower than $\mathrm{COCP}=p_{X,c}$ of Prop. \ref{prop:ocp}. However, the difference is usually small (see examples in Section \ref{s:examples-discussion}) whereas the determination of DOCP is often times quite involved (check the references provided in Section \ref{s:examples-discussion}). Moreover, in case of (b2), DOCP can often be recovered as follows:
\begin{itemize}
    \item take $n_c$ s.t. $(n_c,p_{X,c})$ solves \eqref{e:p_Xc_system};
    \item put $\mathrm{DOCP}=\max(p_{\lfloor n_c \rfloor},p_{\lceil n_c \rceil})$. 
\end{itemize}
In case of (b1), DOCP is very likely to coincide with COCP.
\end{rem}
\begin{rem}\label{r:about_typical_bif_type}
    We are inclined to think that bifurcations (b1)--(b2) are the prevalent ones since we are unaware about practical examples of (b0) satisfying our assumptions. Yet Subsection \ref{ss:p_other} contains an example showing that a counterpart of (b0) may occur on the discrete scale. We were unable to exclude this type theoretically. Thus, appealing to the mentioned example, we are inclined to think that (b0) is not a redundant case but an exceptional one, corresponding to optimal procedures (see the discussion Subsection \ref{ss:discussion}).    
\end{rem}
\begin{rem}\label{r:about_lower_cut_point}
    We did not analyze the behaviour of the lower cut-point $\mathrm{LCP}:=p_l$ of Prop. \ref{prop:bifurcations} because our intent was to consider procedures making sense for all $p$'s in $(0,\mathrm{OCP})$. In certain cases, the behaviour of $p_l$ can be of primary importance. For example, Yao and Hwang \cite{yao_optimal_1990} investigated pairwise BGT algorithm which was known to perform optimally only for $p$'s lying sufficiently far away from origin. Hence, in their case, $p_l$ (not $p_{X,c}$) was the primary quantity of interest.
\end{rem}

\section{Examples and Discussion}\label{s:examples-discussion}

In this section, we provide several examples demonstrating applications of Prop. \ref{prop:bifurcations}. We also provide a couple of examples of the procedures violating our conditions. For the sake of convenience, we put $q:=1-p$ (and parameterize by this parameter as well).

\subsection{Dorfman procedure D}\label{ss:p_dorfman}
We have already described Dorfman procedure in the Introduction \ref{s:intro}. Its description implies that
\begin{gather}\label{e:Dorfman_t_and_M}
    M_D(n,p)=1\cdot q^n + (n+1)(1-q^n)=n+1-nq^n, \\
    t_D(n,p)=\frac{MD(n,p)}{n}=1+\frac{1}{n} - q^n. \nonumber
\end{gather}
We put $c=2$ in (M0). (M1) obviously holds. Since $\frac{\partial}{\partial p}t_D(n,p)=nq^{n-1}>0$ for all $n\in(2,\infty)$, (M2) holds as well. As for (M3), note that
\begin{multline*}
    \frac{\d}{\d n} t_D(n,\mathrm{UCP})=\frac{\d}{\d n}\left(
    1+\frac{1}{n}-\left(\frac{\sqrt{5}-1}{2}\right)^n
    \right)=\\
    -\frac{1}{n^2}-\left(\frac{\sqrt{5}-1}{2}\right)^n\ln \left(\frac{\sqrt{5}-1}{2}\right).
\end{multline*}
Equating this to zero (and solving numerically\footnote{calculations were accomplished by making use of SciPy \cite{scipy} --- a package for Scientific Computing in Python}) one finds out that this function has a unique minimum at $n_{\min}\approx 2.888$ and $\min\limits_{n>2} t_{D}(n,\mathrm{UCP})\approx t_D(2.888,\mathrm{UCP})=1.097$. Moreover, it has maximum at $n_{\max}\approx 5.75$ and then decreases to $\lim\limits_{n\to\infty}t_D(n,\mathrm{UCP})=1$. Finally, from \eqref{e:Dorfman_t_and_M} it follows that \[\forall n\in(2,\infty) \ \lim\limits_{p\to0+}t_D(n,p)=\frac{1}{n}.\] Therefore, (M4) holds as well.

\begin{figure}[h!]
\centering
\includegraphics[width=9cm, height=7cm]{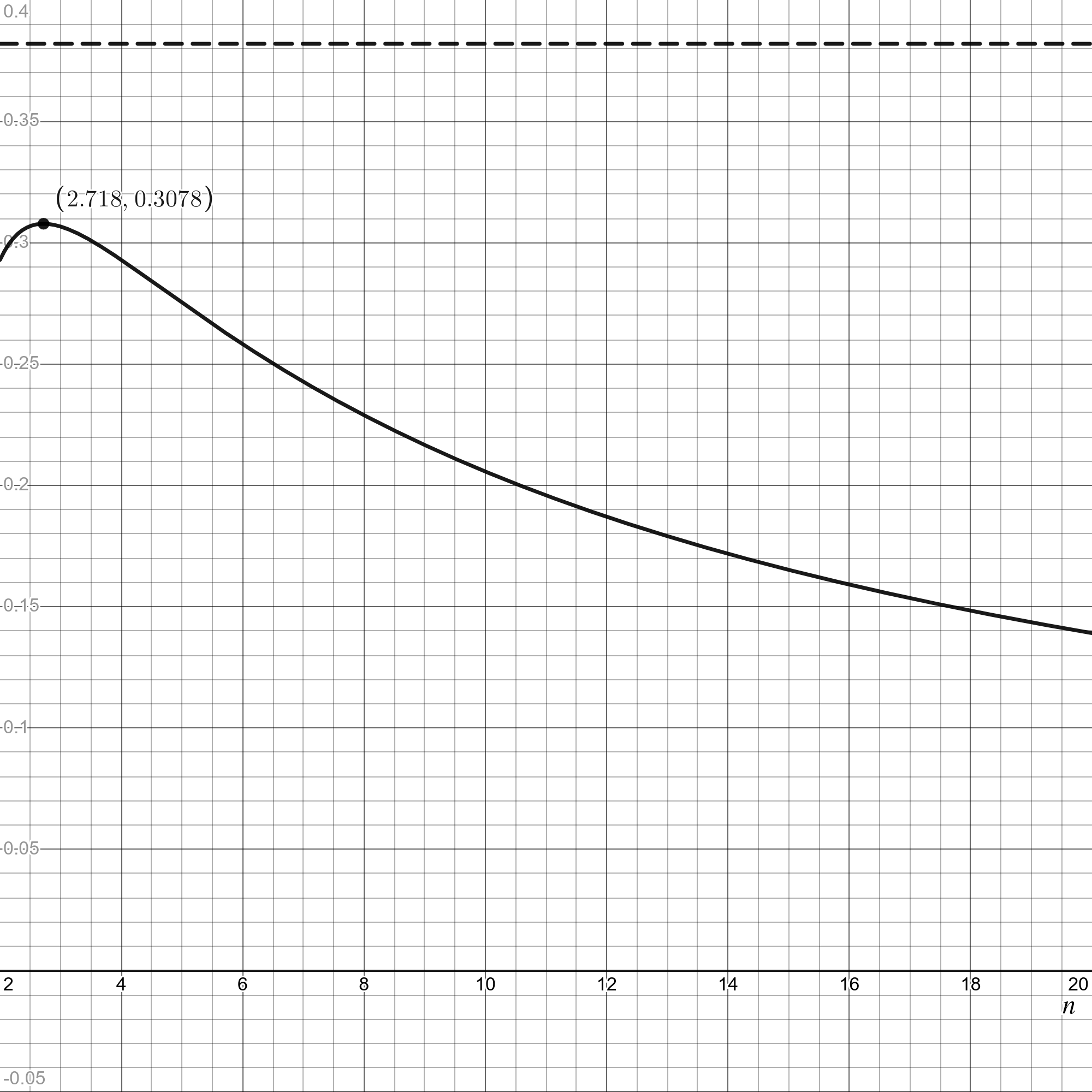}
\caption{plot of $n\mapsto p_n$ (solid line) for the procedure D; dashed line shows constant line $p=\mathrm{UCP}=\frac{3-\sqrt{5}}{2}$; maximal value yields $\mathrm{COCP}=1-\e^{-\e^{-1}}\approx 0.3078$; value $p_{\lceil \e \rceil}=1-3^{-3^{-1}}$ is equal to DOCP.} 
\label{fig:Dorfman_p_n}
\end{figure}

Figure \ref{fig:Dorfman_p_n} shows a plot of the inverted bifurcation map $n\mapsto p_n$ described in Prop. \ref{prop:bifurcations}. In this case, it admits analytical expression: $p_n = 1-\left(\frac{1}{n}\right)^{\frac{1}{n}}$. System \eqref{e:p_Xc_system} is given by
\begin{equation}
    \begin{cases}
        &\frac{1}{n}=q^n,\\
        &-\frac{1}{n^2}=q^n\ln q,
    \end{cases}
\end{equation}
and can be solved analitically too. Its solution is $(n*,p_{D,c})=(\e,1-\e^{-\e^{-1}})$. Samuels \cite{samuels_exact_1978} investigated procedure D on the discrete scale. His analysis revealed that $\mathrm{DOCP}=1-3^{-3^{-1}}$. Our COCP is quite close. Moreover, we can recover DOCP by applying the method described in the Remark \ref{r:about_scale}. In this particular case the method works and affirms Samuel's result.

\subsection{Squared Array procedure A2}\label{ss:p_A2}

Squared Array (A2) procedure was introduced by Phatarfod and Sudbury \cite{Phatarfod-1994} (and later generalized by Berger, Mandell and Subrahmanya \cite{Berger-2000}). To apply A2, one places $n^2,n\in\{2,3,\ldots\}$ items on $n\times n$ matrix and applies the following tests:
\begin{itemize}
    \item $n$ tests of pooled samples corresponding to row pools;
    \item $n$ tests of pooled samples corresponding to column pools;
    \item individual tests on items having positive both row and column tests.
\end{itemize}
This way all defectives get identified. 

It was shown by Phatarfod and Sudbury \cite{Phatarfod-1994} that
\begin{equation}\label{e:mean_T_N_A2}
    M_{A2}(n,p)=2n + n^2\left(1-2q^n + q^{2n-1}\right).
\end{equation} 
Noting that the tested cohort has $N(n)=n^2$ items in total, we therefore have that
\begin{equation}\label{e:t_N_A2}
    t_{A2}(n,p)=\frac{2}{n} + 1-2q^n + q^{2n-1}=\frac{2}{n} + (1-q^n)^2+pq^{2n-1}.
\end{equation} 
From the latter expression, it follows that A2 makes sense only for $n>2$. Bearing in mind practical aspect (i.e., the fact that cohort sizes are integers) we therefore set $c=3$ in (M0). Note that, due to the design of the procedure, this truncation actually restricts sizes of the tested cohorts to start from $9=3\times 3$ (not $3$).  It is obvious that (M1) holds. Since $\forall p\in (0,\mathrm{UCP}]$
\begin{multline*}
    \frac{\partial}{\partial p} M_{A2}(n,p)=n^2\left(2nq^{n-1}-(2n-1)q^{2n-2}\right)=\\
    n^2\left(q^{2n-2} + 2nq^{n-1}(1-q^{n-1})\right)>0,
\end{multline*}
(M2) holds as well.

Justification of (M3) can be done by accomplishing the following steps:
\begin{itemize}
    \item check that $\frac{\partial}{\partial n}t_{A2}(n,p)=-\frac{2}{n^2}-2q^{n}\ln q(1-q^{n-1})$ and $\frac{\partial^2}{\partial n^2}t_{A2}(n,p)=\frac{4}{n^3}-2q^{n}\ln^2 q(1-2q^{n-1})$;
    \item numerically solve $\frac{\partial^2}{\partial n^2}t_{A2}(n,p)\Big|_{p=\mathrm{UCP}}=0$ and obtain two roots: $n_1\approx 5.278,n_2\approx 9.448$;
    \item verify that $n_1$ corresponds to maximum whereas $n_2$ corresponds to minimum of $n\mapsto \frac{\partial}{\partial n}t_{A2}(n,\mathrm{UCP})$ and that $\frac{\partial}{\partial n}t_{A2}(n_1,\mathrm{UCP})<-0.0055<0$;
    \item conclude that $n\mapsto t_{A2}(n,\mathrm{UCP})$ is decreasing and (M3) holds since \[\lim\limits_{n\to\infty}t_{A2}(n,\mathrm{UCP})=1.\]
\end{itemize}

\begin{figure}[h!]
\centering
\includegraphics[width=9cm, height=7cm]{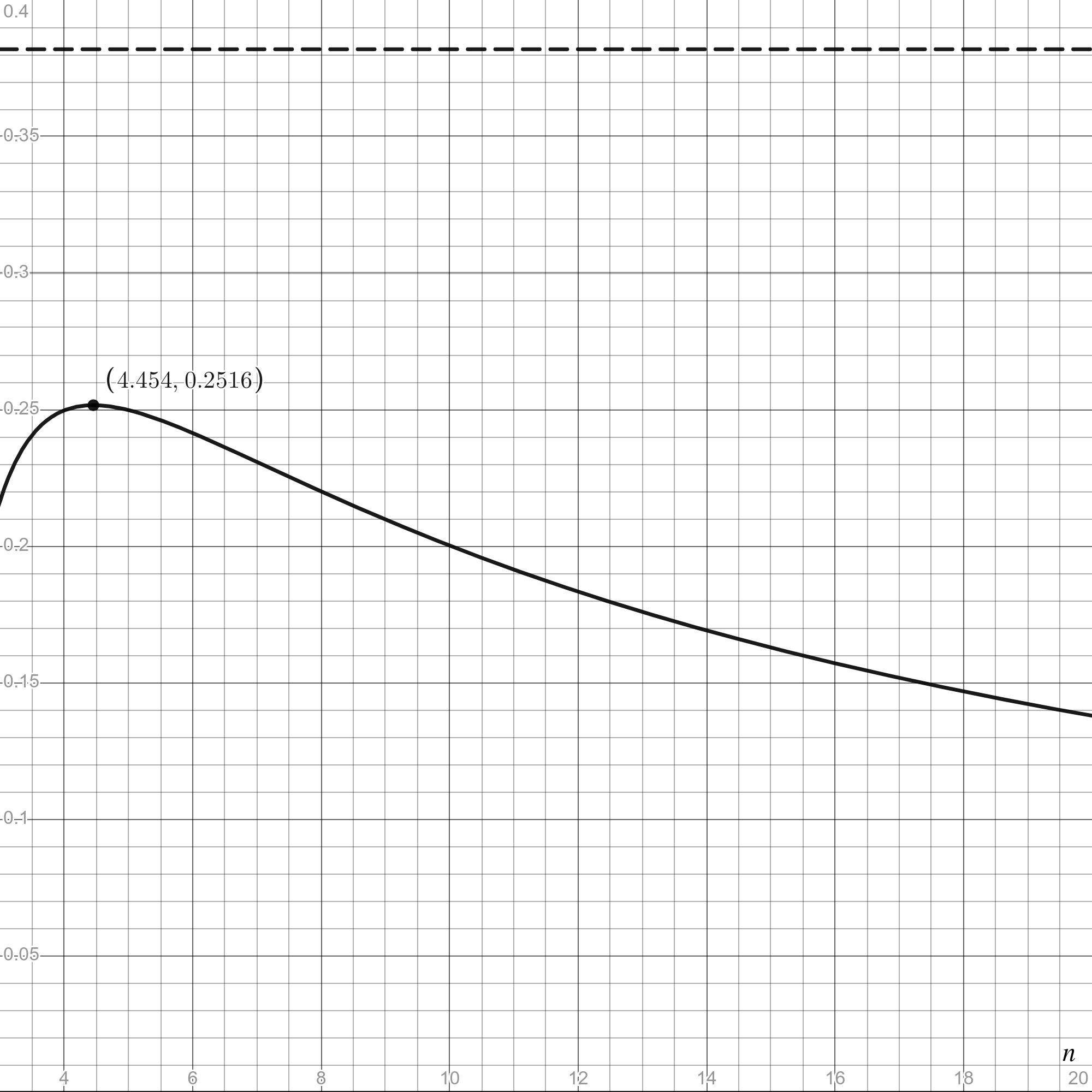}
\caption{plot of $n\mapsto p_n$ (solid line) for the procedure A2; dashed line shows constant line $p=\mathrm{UCP}=\frac{3-\sqrt{5}}{2}$; maximal value yields $\mathrm{COCP}\approx 0.2516$; $p_{5}\approx0.2498$ is equal to DOCP.} 
\label{fig:A2_p_n}
\end{figure}

Figure \ref{fig:A2_p_n} shows the inverted bifurcation curve $(3,\infty)\ni n\mapsto p_n$ of Prop. \ref{prop:bifurcations}. System \eqref{e:p_Xc_system} is given by 
\begin{equation*}
    \begin{cases}
        &\frac{2}{n}-2q^n+q^{2n-1}=0,\\
        &-\frac{1}{n^2}-q^n(1-q^{n-1})\ln q=0.
    \end{cases}
\end{equation*}
As can be seen from the curve, it has a unique solution $(n^*,p_{A2,c})\approx (4.454,0.252)$. Kim and Hudgens \cite{hudgens_optimal_2011}  have investigated A2 on the discrete scale. They have proved that $\mathrm{DOCP}=0.2498$. This point precisely coincides with $p_{\lceil n^*\rceil}$. Thus, the method described in \ref{r:about_scale} again led to the recovery of the DOCP.

\subsection{Modified Dorfman procedure MD}\label{ss:p_m_dorfman}

\begin{figure}[h!]
\centering
\includegraphics[width=9cm, height=7cm]{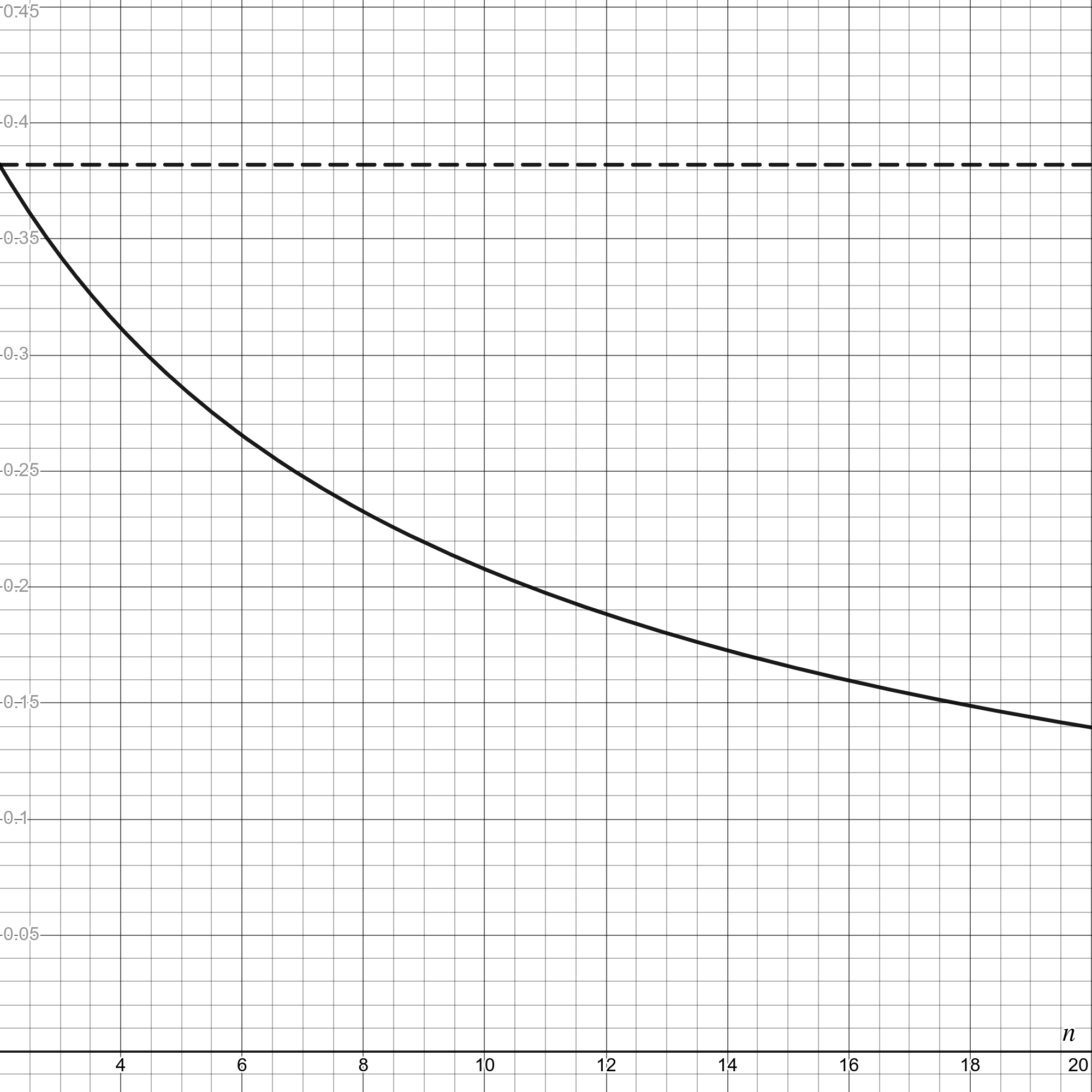}
\caption{plot of $n\mapsto p_n$ (solid line) for the procedure MD; dashed line shows constant line $p=\mathrm{UCP}=\frac{3-\sqrt{5}}{2}$.} 
\label{fig:mDorfman_p_n}
\end{figure}

The original Dorfman procedure D is inconsistent: if the pooled sample tests positive and all but the last individual items test negative, it still tests the last one. In such case, Sobel and Groll \cite{sobel+groll:1959}  suggested not to test the last item since test result can be inferred. We call their procedure Modified Dorfman (MD). For MD,
\begin{equation}\label{e:t_MD}
    t_{MD}(n,p)= 1- q^{n}+\frac{1-pq^{n-1} }{n}.
\end{equation}
We set $c=2$ and (M0) is satisfied. As usually, (M1) is obvious. Since 
\begin{multline*}
    \frac{\partial}{\partial p} t_{MD}(n,p)=nq^{n-1}-\frac{1}{n}\left(q^{n-1}-(n-1)pq^{n-2}\right)=\\
    q^{n-1}\left(n-\frac{1}{n}\right) + \frac{n-1}{n}pq^{n-2}>0
\end{multline*}
for any fixed $n\in(2,\infty)$, (M2) holds. $\forall n\in(2,\infty) \lim\limits_{p\to 0+}t_{MD}(n,p)=\frac{1}{n}<1$. Hence (M4). Finally, verification of (M3) can be done in the same as in case of procedure A2. Since exercise is quite lengthy and tedious, we omit the details as well as checking that system \eqref{e:p_Xc_system} does not admit solution lying in $(2,\infty)\times (0,\mathrm{UCP}]$. The latter means that we have bifurcation of type (b1). Since $t_{MD}(2,\mathrm{UCP)=1}$, we conclude that $p_{MD,c}=\lim\limits_{n\to 2+}p_n=\mathrm{UCP}$. Figure \ref{fig:mDorfman_p_n} provides graphical illustration of the said.

\subsection{Sterrett procedure S}\label{ss:p_sterrett}
\begin{figure}[h!]
\centering
\includegraphics[width=9cm, height=7cm]{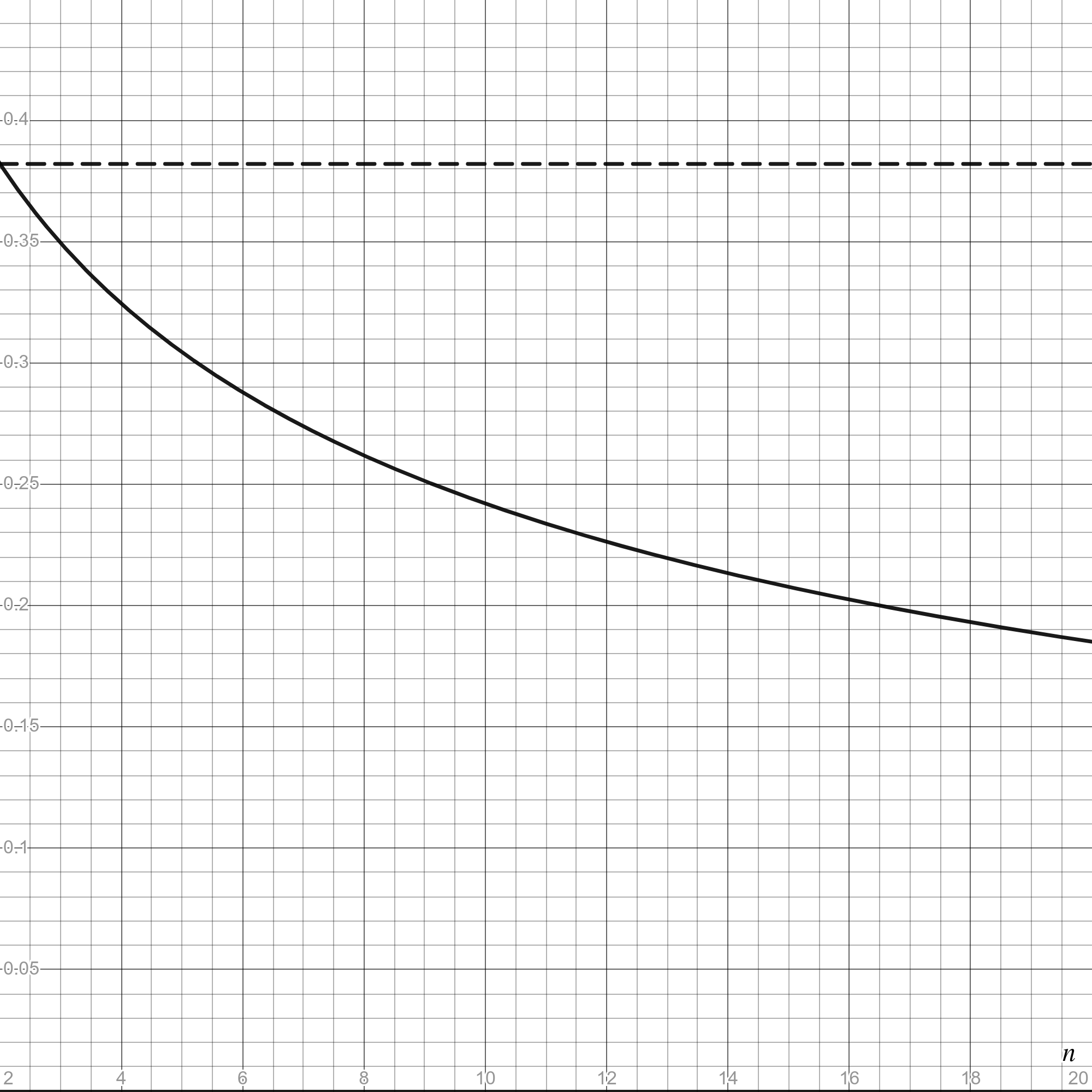}
\caption{plot of $n\mapsto p_n$ (solid line) for the procedure S; dashed line shows constant line $p=\mathrm{UCP}=\frac{3-\sqrt{5}}{2}$.} 
\label{fig:Sterrett_p_n}
\end{figure}

Sterrett \cite{sterrett_detection_1957} suggested to modify Dorfman's procedure as follows:
\begin{itemize}
    \item[(s1)] test pool consisting of all items; in case it tests negative, finish; otherwise proceed to the step (s2).
    \item[(s2)] test items one-by-one until the first positive appears; consider the remaining untested set as an initial one and go back to the step (s1).
\end{itemize}
Sobel and Groll \cite{sobel+groll:1959} demonstrated that
\begin{equation}\label{e:tS_n}
    t_{S}(n,p)= 2 - q + \frac{2q-(1-q)^{-1}(1-q^{n+1})}{n}.
\end{equation}
We set $c=2$ in (M0). (M1) then readily holds. An equivalent way to verify (M2) lies in showing that $(1-\mathrm{UCP},1)\ni q\mapsto t_S(n,p)$ is decreasing. Since
\begin{equation*}
    \frac{\partial}{\partial q} t_S(n,p)=\left(\frac{2}{n}-1\right)-\frac{1}{n}\left(\frac{1-q^{n+1} }{1-q}\right)_q^{\prime},
\end{equation*}
one sees that the term $2/n-1<0$ for any $n>2$ and it suffices to show that the derivative 
\begin{equation*}
    \left(\frac{1-q^{n+1} }{1-q}\right)_q^{\prime}=
    \frac{1-(n+1)q^{n}+nq^{n+1}}{(1-q)^2}
\end{equation*}
is positive. We omit the details of this exercise. (M4) follows by noting that, for any fixed $n>2$,
\begin{equation*}
    \lim_{p\to 0+}t_{S}(n,p)= \lim_{q\to 1-}t_{S}(n,p)
    =1+\frac{2-\lim\limits_{q\to1-}\frac{\left(1-q^{n+1} \right)_q^{\prime}}{\left(1-q\right)_q^{\prime}}}{n}
    =1+\frac{2-\frac{n+1}{1}}{n}=\frac{1}{n}<1.
\end{equation*}
As in the previous example, we omit verification of (M3). It amounts to careful analysis of the derivative of $(2,\infty)\ni n\mapsto t_{S}(n,\mathrm{UCP})$. One can also show that the system \eqref{e:p_Xc_system} does not admit solutions $(n,p)$ lying in $(2,\infty)\times(0,\mathrm{UCP}]$. Since $\mathrm{UCP}$ solves $t_{S}(2,p)=1$ (w.r.t. $p$), we again have that $p_{S,c}=\lim_{n\to 2+} p_n=\mathrm{UCP}$ as in the previous example. Figure \ref{fig:Sterrett_p_n} demonstrates that the bifurcation curve qualitatively exhibits the same behaviour too.

\subsection{Examples violating our assumptions}\label{ss:p_other}

We have already mentioned the work by Yao and Hwang \cite{yao_optimal_1990} in Remark \ref{r:about_lower_cut_point}. Without digging into details, the pairwise testing (PT) procedure investigated there has a mean
\begin{equation*}
    M_{PT}(n,p)=n\frac{2-q^2}{1+q}+\frac{q^2+q-1}{(1+q)^2}(1-(-q)^n).
\end{equation*}
Clearly, it can not be extended to the continuously differentiable function. Hence, (M1) does not hold. (M3)
does not hold as well. Indeed, $\mathrm{UCP}$ solves $q^2+q-1=0$; also $\frac{2-q^2}{1+q}\Big|_{p=\mathrm{UCP}}=1$ and 
\begin{equation}\label{e:pt_is_b0}
\frac{2-q^2}{1+q}<1 \Longleftrightarrow 0<q^2+q-1 \Longleftrightarrow p\in (0,\mathrm{UCP}).    
\end{equation}
Another example of this kind is the Halving procedure (H) suitable for testing cohort having $2^n,n\in\Nd$ items. Its mean
\begin{equation*}
    M_{H}(n,p)=1+2^{n+1}\sum_{k=1}^n \frac{1-q^{2^k}}{2^k} 
\end{equation*}
again violates (M1). We were unable to trace back the roots of this procedure with certainty yet Johnson \cite{johnson_inspection_1991} provides a full chapter (see Chapter 10) devoted to an in-depth analysis of its generalization allowing for an imperfect testing.

\subsection{Discussion}\label{ss:discussion}

The main limitation of our method is the assumption (M1). All other assumptions are much more likely to be true when talking about a typical GT procedure. At a first glance, it may seem that (M3) rules out procedures having $p_{X,c}=\mathrm{UCP}$. As demonstrated by example, this is not the case. In fact, it restricts the subset of GT procedures to those having $p_{X,c}$ on the boundary of the domain of the bifurcation curve. We are inclined to think that this way we rule out optimal procedures, i.e. those which are best performing in certain classes. However, we do not treat this as a drawback since for optimal procedures one generally expects that $p_{X,c}=\mathrm{UCP}$. Expanding in this direction, we note that relationships \eqref{e:pt_is_b0} uncover an interesting fact mentioned previously: \[\forall n\in(2,\infty)\,  t_{PT}(n,\mathrm{UCP})=1\] which means that PT procedure is of type (b0). Yao and Hwang \cite{yao_optimal_1990} have proved that PT is an optimal nested testing procedure\footnote{we do not define the class of the nested procedures here and refer an interested reader to \cite{yao_optimal_1990}; for $p=\mathrm{UCP}$, they treat PT procedure as an optimal one; this is a matter of opinion since $t_{PT}(n,\mathrm{UCP})=1$ and the same result holds for one-by-one testing} if and only if $p\in\left[1-\frac{\sqrt{2}}{2},\frac{3-\sqrt{5}}{2}\right]$. This suggests that procedures of type (b0) are likely to be those which are optimal in some region.

\begin{figure}[h!]
\centering
\includegraphics[width=9cm, height=7cm]{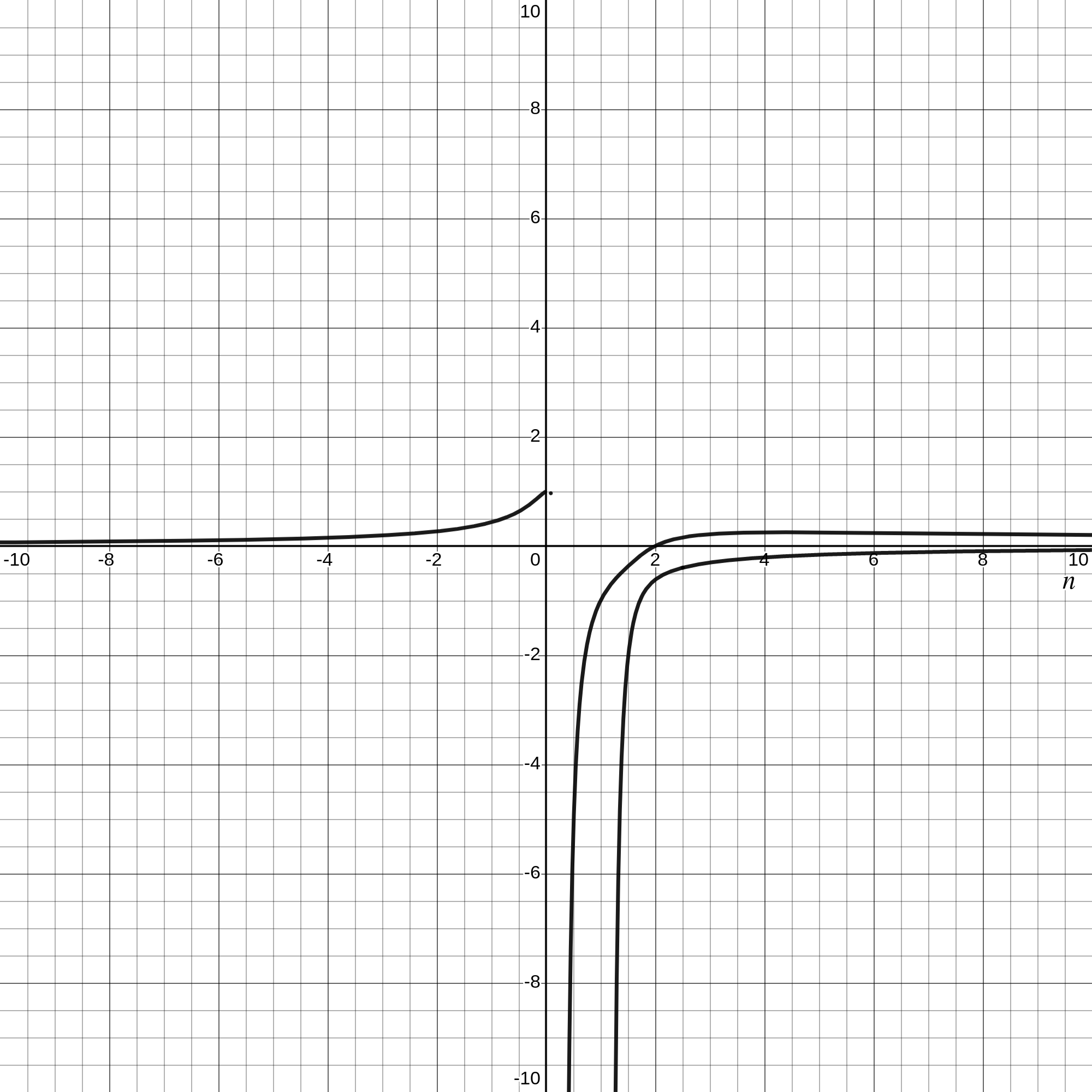}
\caption{plot of bifurcation curve for procedure A2 when the domain of the dynamical system \eqref{e:ds1} is extended to the whole real line.} 
\label{fig:A2_p_n_extended}
\end{figure}

In case (M0)--(M4) hold, our algorithm appears to be efficient. There is a word of caution: one has to choose $c$ in (M0) carefully. The point is that the dynamical system \eqref{e:ds1}, when viewed on a wider domain, may exhibit more complicated bifurcations. Figure \ref{fig:A2_p_n_extended} provides a convincing graphical illustration. Turning to the types of bifurcations, one sees that, in terms suggested by Strogatz \cite{strogatz_nonlinear_2015}, (b1) is usually the saddle point bifurcation, whereas for (b2) the system admits fixed points for all but boundary value of the control parameter $p\in(0,\mathrm{UCP}]$. We are inclined to think that the dynamical system approach taken by us could be extended to GT procedures violating (M1) and successfully used to investigate other general properties of GT procedures yet one needs to work on the discrete scale.

\appendix

\section{Proofs}

\emph{Proof of Proposition \ref{prop:ocp}}. 

Take $p\in(0,p_{X,c})$. By definition of $p_{X,c}$, there exists $n\in(c,\infty): t_X(n,p)<1$. Hence, $X$makes sense for that $p$. Since $p\in(0,p_{X,c})$ was arbitrary, it holds true for all $p\in(0,p_{X,c})$.

Next, assume that $p_{X,c}<\mathrm{UCP}$ (otherwise implication \eqref{e:makes_no_sense} is obvious) and take $p\in(p_{X,c},\mathrm{UCP}]$. Case "$\exists n\in(c,\infty):t_X(n,p)<1$" contradicts the definition of $p_{X,c}$. Hence, $\forall n\in(c,\infty)\ t_X(n,p)\geq 1$. Assuming that $t_X(n,p)=1$ for some $n\in(c,\infty)$ again leads to contradiction. Indeed, take $p^\prime\in(p_{X,c},p)$ and employ (M2) to deduce that 
\begin{multline}\label{e:contradiction1}
    t_X(n,p^\prime)<t_X(n,p)=1\imply p_{X,c}=\\
    \sup\{p\in(0,\mathrm{UCP})\mid \exists n\in(c,\infty): t_X(n,p)<1\}\geq 
    p^\prime> p_{X,c}. \qed
\end{multline}

\emph{Proof of Proposition \ref{prop:bifurcations}}. First, note that $t_X(n,p_{X,c})\geq 1$ for all $n\in(c,\infty)$ provided $p_{X,c}<\mathrm{UCP}$. Assuming an opposite leads to the same contradiction as in \eqref{e:contradiction1}. With this in view, we proceed to the analysis of the distinct types of possible bifurcations.

\smallskip\emph{Case (b0)}. Since $p_{X,c}$ solves $t_X(n,p_{X,c})=1$ for some $n\in(c,\infty)$, (M3) implies that $p_{X,c}<\mathrm{UCP}$. Further, note that 
\begin{equation}\label{e:caseb0_a1}
    \forall (n,p)\in (c,\infty)\times(0,p_{X,c})\ t_X(n,p)<1.
\end{equation}
must hold since existence of $(n,p)\in(c,\infty)\times (0,p_{X,c})$ s.t. $t_X(n,p)\geq 1$ contradicts the premise\footnote{$t_X(n,p)= 1$ is clearly impossible; assuming strict inequality $t_X(n,p)> 1$, to reach contradiction, employ (M2) and (M4)} "$p_{X,c}$ is the only value of the control parameter for which dynamical system \eqref{e:p_Xc_system} admits fixed points in $(c,\infty)$".

Fix arbitrary $(n_1,p_1)\in(c,\infty)\times(0,p_{X,c})$ and take $p_2\in(p_{X,c},\mathrm{UCP})$. By Prop. \ref{prop:ocp}, $t_X(n_1,p_2)>1$. By (M2), $[p_1,p_2]\ni p\mapsto t_X(n_1,p)$ is strictly increasing. Therefore, there exists unique $p_0\in(p_1,p_2)$ s.t. $t_X(n_1,p_0)=1$. By the premise, $p_0=p_{X,c}$. Since $n_1$ was arbitrary, it follows that 
\begin{equation}\label{e:caseb0_key_property}
    \forall\,n\in(c,\infty)\,t_X(n,p_{X,c})=1
\end{equation}
and $p_{X,c}$ is the unique value of the control parameter $p\in(0,\mathrm{UCP})$ obeying this property. Differentiating both sides of \eqref{e:caseb0_key_property} with respect to $n$ one finds out that \eqref{e:p_Xc_system} holds as well.

\smallskip\emph{Case (b1)--(b2)}. Assume that there exists $(n_l,p_l)\in(c,\infty)\times(0,p_{X,c})$ s.t. $t_X(n_l,p_l)=1$. Take arbitrary $p\in(p_l,p_{X,c})$. Since $p>p_l$, it follows that $t_X(n_l,p)>1$ because of (M2). On the other hand, by the definition of $p_{X,c}$, there exists $n_p\in(c,\infty)$ s.t. $t_X(n_p,p)<1$. Therefore, by the Intermediate Value Theorem and continuity of $n\mapsto t_X(n,p)$, there exists $n_1\in (\min(n_p,n_l),\max(n_p,n_l))$ s.t. $t_X(n_1,p)=1$. Since this holds for any $p\in[p_l,p_{X,c})$, we have (b1) provided $\forall n\in (c,\infty)\, t_X(n,p_{X,c})>1$. Otherwise, we have (b2), and $p_{X,c}$ then can't be equal to $\mathrm{UCP}$ because of (M3).

\smallskip\emph{Inversion of the bifurcation curve}. When dealing with (b0), we have already shown that the bifurcation curve defines a constant map $(c,\infty)\ni n \mapsto p_n\equiv p_{X,c}$. As for (b1)--(b2), note that, for any fixed $n\in(c,\infty)$, the following applies:
\begin{itemize}
    \item by the said in the very beginning of the proof and (M3),
    \begin{equation}\label{e:bif_starting_remark}
        \forall\, n\ t_X(n,p_{X,c})\geq 1;
    \end{equation}
    \item by (M4),
    \begin{equation}\label{e:bif_M4}
        \forall n\,\exists p\in(0,p_{X,c}): t_X(n,p)<1;
    \end{equation}
    \item \eqref{e:bif_starting_remark}--\eqref{e:bif_M4} and (M2) imply existence of a unique $p_n\in(0,p_{X,c}]$ s.t. $t_X(n,p_n)=1$.
\end{itemize}
Therefore, we have a well defined map $(c,\infty)\ni n\mapsto p_n\in(0,p_{X,c}]$. By (M1)--(M2), $p\mapsto t_X(n,p)$ is differentiable and increasing for any $n\in(c,\infty)$. Therefore, $\forall n\in (c,\infty)\ \frac{\partial}{\partial p}t_X(n,p)>0$ and one can apply the Implicit Function Theorem to $\fy(n,p)=t_X(n,p)-1$ to deduce that $n\mapsto p_n$  is differentiable as well. Moreover, differentiating both sides of $t_X(n,p_n)=1$ and applying the chain rule, we have that
\begin{equation}\label{e:derivative_of_p_n}
    \frac{\partial}{\partial n}t_X(n,p_n) + \frac{\partial}{\partial p}t_X(n,p_n)\frac{\partial}{\partial n}p_n=0\imply
    \frac{\partial}{\partial n}p_n = -\frac{\frac{\partial}{\partial n}t_X(n,p_n)}{\frac{\partial}{\partial p}t_X(n,p_n)}.
\end{equation}
Therefore, looking for extremes of $n\mapsto p_n$ one has to solve \[\frac{\partial}{\partial n}t_X(n,p_n)=0 \Longleftrightarrow \frac{\partial}{\partial n}p_n=0\] with respect to $n$. Since $p_n$ also solves $t_X(n,p)=1$, extremes and corresponding values can be obtained by solving \eqref{e:p_Xc_system}. For bifurcations of type (b0) and (b2), maximal value $p_{X,c}$ is attained at some inner point(s) $n_c\in(c,\infty)$; for the bifurcation of type (b1), the maximal value lies on the boundary of its domain. $\qed$

\bibliographystyle{plain}

\section{Figures}
The figures were produced by making use of Desmos Graphing Calculator \cite{desmos_graphing_calc}. 

\end{document}